# Surface versus Bulk Characterization of the Electronic Inhomogeneity in a VO₂ Film


Y. J. Chang, J. S. Yang, Y. S. Kim, D. H. Kim, and T. W. Noh[*]

*ReCOE & FPRD, Department of Physics and Astronomy, Seoul National University, Seoul 151-747, Korea*

## D.-W. Kim

*Department of Applied Physics, Hanyang University, Ansan, Kyeonggi 426-791, Korea*

## E. Oh, and B. Kahng

*CTP & FPRD, Department of Physics and Astronomy, Seoul National University, Seoul 151-747, Korea*

## J.-S. Chung

*Department of Physics and CAMDRC, Soongsil University, Seoul 156-743, Korea*



We investigated the inhomogeneous electronic properties at the surface and interior of VO₂ thin films that exhibit a strong first-order metal-insulator transition (MIT). Using the crystal structural change that accompanies a VO₂ MIT, we used bulk-sensitive X-ray diffraction (XRD) measurements to estimate the fraction of metallic volume $p^{\mathrm{XRD}}$ in our VO₂ film. The temperature dependence of the $p^{\mathrm{XRD}}$ was very closely correlated with the $dc$ conductivity near the MIT temperature, and fit the percolation theory predictions quite well: $\sigma \sim (p - p_{\mathrm{c}})^t$ with $t = 2.0 \pm 0.1$ and $p_{\mathrm{c}} = 0.16 \pm 0.01$. This agreement demonstrates that in our VO₂ thin film, the MIT should occur during the percolation process. We also used surface-sensitive scanning tunneling spectroscopy (STS) to investigate the microscopic evolution of the MIT near the surface. Similar to the XRD results, STS maps revealed a systematic decrease in




the metallic phase as temperature decreased. However, this rate of change was much slower than the rate observed with XRD, indicating that the electronic inhomogeneity near the surface differs greatly from that inside the film. We investigated several possible origins of this discrepancy, and postulated that the variety in the strain states near the surface plays an important role in the broad MIT observed using STS. We also explored the possible involvement of such strain effects in other correlated electron oxide systems with strong electron-lattice interactions.

# I. INTRODUCTION

The unusual physical properties of transition metal oxides (TMOs) have made them one of the most widely investigated materials in condensed matter physics over the last two decades. These properties include a metal-insulator transition (MIT),[1, 2] colossal magnetoresistance (CMR),[3-6] and high temperature superconductivity.[7-10] The strong electron correlation, represented by the Hubbard Hamiltonian, has provided valuable insight into the intriguing behavior of TMOs, including their electronic structures. Researchers have used numerous experimental tools, including optical, photoelectron (PES), and scanning tunneling (STS) spectroscopies, to investigate the evolution of the electronic structural changes as electron correlation increases systematically.

Recently, numerous reports have examined the electronic inhomogeneities in TMO materials. For example, atomic resolution STS studies of the high-temperature superconductor $Bi_2Sr_2CaCu_2O_y$ revealed intrinsic electronic inhomogeneities that might be



related to basic mechanisms of the superconducting transition.[7-10] The investigation of such intrinsic inhomogeneities requires obtaining atomically clean surfaces by cleaving single crystal samples. This cleaving process is relatively easy for layered oxides. However, techniques for preparing an atomically well-defined surface of non-layered oxides are not well established or are challenging, making it difficult to obtain atomically flat surfaces. Therefore, many experimental works have investigated electronic inhomogeneities using non-cleaved samples or thin films.[3-6] The results of experiments conducted under these conditions can vary widely depending on the probing depth of the experimental tools used. These differences can be very significant, particularly for some TMO systems with strong electron-lattice interactions.

Vanadium dioxide ($VO_2$) is one of these widely investigated TMOs with a strong electron correlation. This compound exhibits an MIT at 341 K and has quite strong electron-lattice couplings.[2, 11-19] Since its MIT is of the first order, it should also exhibit large related electronic inhomogeneities. Optical spectra of a $VO_2$ film suggested the coexistence of metallic and insulating phases during its MIT, the evolution of which could be explained in terms of the percolation model.[13] Our recent and preliminary STS works on another $VO_2$ film suggested similar phase segregation phenomena, but its temperature-dependent evolution appeared to be much slower than the optical data.[14] These studies raised the question do electronic inhomogeneities at the $VO_2$ surface behave differently from those inside the bulk?

In this study, we investigated the phase transition of a $VO_2$ thin film as it underwent MIT, using both surface- and bulk-sensitive experimental tools. STS studies provided



microscopic visual images of surface inhomogeneities, while X-ray diffraction (XRD) patterns monitored the structural phase transition and demonstrated the coexistence of phases with two differing crystal structures, possibly metallic and insulating phases, inside our $VO_2$ film. Both these surface- and bulk-sensitive results indicated the presence of electronic phase separation in the $VO_2$ film, but this differed in two important ways. First, the surface underwent transition over a much wider range of temperature (300–365 K) than did the bulk (330–340 K). Second, the fraction of metallic volume, obtained from XRD data, well reproduced *dc* conductivity ($\sigma$) based on the percolation model; however it was not possible to fit the $\sigma$ data reasonably using STS data.

Section II of this paper describes our experimental methods. Section III presents the experimental results for *dc* transport, XRD, and STS measurements. It also explains how we were able to analyze our STS maps and XRD spectra to obtain quantitative values for the area and volume fractions of the metallic phase, respectively. Section IV applies the percolation model to assess whether the values we obtained for the fraction of the metallic volume represent an MIT occurring inside the $VO_2$ film. It also compares our experimental results with results from earlier optical and photoemission spectroscopy studies. When we compared our experimental results with experiments using different probing depths, we found that the temperature dependence of the MIT inside our $VO_2$ film differed considerably from that at the surface. Based on our findings, we revisited some earlier studies including PES and STS research on TMO materials. Section V summarizes the main findings of this study.



## II. EXPERIMENTAL METHODS

We successfully grew epitaxial $VO_2$ thin films on $Al_2O_3$ (012) single crystal substrates using a pulsed laser deposition technique.[14] We prepared a sintered polycrystalline $V_2O_3$ target using the conventional solid-state reaction method. We focused a KrF excimer laser beam on the target with a fluence of 1 J/cm$^2$ and a repetition rate of 2 Hz. To obtain high equality epitaxial films, we used a relatively slow growth rate of 0.4 nm/min. To obtain a sharp MIT transition, we optimized the film deposition conditions using *dc* resistance data. We found the optimal substrate temperature and oxygen pressure were 450 °C and 15 mTorr, respectively. After deposition, we cooled the films to room temperature at the deposition oxygen pressure without any post-annealing process. Subsequent XRD studies confirmed the growth of high-quality epitaxial $VO_2$ thin films without any evidence of impurity phases.[14]

After growing the 70-nm $VO_2$ film, we quickly transferred it into an ultrahigh vacuum (UHV) chamber (~ 5×10$^{-11}$ Torr) and conducted STM and STS studies. Our UHV chamber was equipped with reflection high-energy electron diffraction (RHEED) and a variable-temperature STM system that could operate from 25–1500 K. Immediately after the transfer, we obtained a spotty RHEED pattern, confirming that our $VO_2$ film surface had good crystallinity. Using a Pt-Ir tip, we obtained topographic images at the sample bias voltage of 2 V, and then carried out STM and STS measurements at various sample bias voltages ranging from –1.0 to 2.0 V. With a fixed tip-sample distance, we took the STS *I-V* curves and differentiated them numerically to obtain *dI/dV-V* curves.



To determine the temperature-dependent evolution of microscopic inhomogeneities related to the MIT, we performed our STM/STS studies using temperatures between 300–380 K. To ensure that the temperature measurements were accurate, we used silicon-diode temperature sensors and calibrated the sample temperature within an experimental error of less than 0.1 K. We increased the sample temperature using a resistive heater, waited for a sufficient time, and then moved the STM tip toward the $VO_2$ film again to adjust for changes caused by thermal drift. The thermal drift was so great that it was impossible to perform a series of STS studies at the same sample position. However, we repeated STS studies several times at each temperature, and were able to reproduce most experimental results.

After the STM/STS studies, we carried out *dc* transport and temperature dependent XRD studies. Using an XRD machine equipped with a heating stage, we conducted $\theta$–$2\theta$ scans near the $VO_2$ (200) peak at temperatures between 320 and 350 K with a 1-K step in ambient atmosphere. To examine the thermal expansion of the substrate, we also measured the changes in the substrate peak, *i.e.* $Al_2O_3$ (012). We found that the substrate had a negligibly small thermal expansion of $2\times10^{-4}$ deg./K, so in our XRD analysis we ignored the contribution made by thermal expansion of the substrate.

### III. RESULTS AND ANALYSIS

#### A. DC-transport properties



Our VO$_2$ film exhibited a clear first-order MIT in a temperature region from 330–340 K. Figure 1 presents $dc$ conductivity ($\sigma$) data for the VO$_2$ thin film, measured using the conventional four-probe $dc$ technique. As temperature increased, the film underwent an MIT from an insulating state to a metallic state at a MIT temperature ($T_{MI}$) of 340 K, a value similar to that for a single crystal, $i.e.$, about 341 K.[11] The MIT of our film was accompanied by a large $dc$ conductivity change ($\Delta\sigma/\sigma_I$) with a fourth-order difference of magnitude, which is comparable to the value for a single crystal, $i.e.$, $\Delta\sigma/\sigma_I \sim 10^5$.[11] Due to the nature of the first-order phase transition, the experimental values of $T_{MI}$ differed during the heating and cooling cycles. The $T_{MI}$ values of our film had a hysteretic difference of about 6.5 K, which was greater than the value of about 2.0 K for a single crystal.[11] The MIT was completed in the temperature region of about 2.2 K. This value of MIT width, $\Delta T_{MI}$, is also greater than that for a single crystal, $i.e.$, <1.0 K.[12]

We obtained some insights into the quality of our VO$_2$ films by comparing our results with those from earlier studies of VO$_2$ thin films. Recently, Nagashima $et$ $al$. reported that the ambient atmosphere affected MIT in strained VO$_2$ ultrathin films grown epitaxially on TiO$_2$ (001) single crystal substrates.[15] By changing the ambient atmosphere pressure, they were able to vary the $\Delta\sigma/\sigma_I$ values of their films from $10^0$ to $10^3$. Films deposited at lower oxygen pressures had smaller $\Delta\sigma/\sigma_I$ values, possibly caused by oxygen vacancies. The large $\Delta\sigma/\sigma_I$ value (about $10^4$) of our film suggests that it is nearly free from such oxygen vacancy effects. Muraoka and Hiroi were able to change the $T_{MI}$ value of their epitaxial films from 341 to 300 K by controlling the strain state.[16] Later, Nagashima $et$ $al$. found that such strain effects could be relieved by developing nanoscale line cracks under thermal stress.[17] The similarity of the $T_{MI}$ value of our film to values for a single crystal suggests that strain effects and nanoscale line cracks do not play important roles in the transport properties of our film. Brassard $et$ $al$. also reported the MIT characteristics of sputtered VO$_2$ films;[18] they found that $\Delta\sigma/\sigma_I$ could vary from $10^2$ to $10^3$, and that $\Delta T_{MI}$ could vary from 3.0 to 9.5 K. They argued that such variation in the MIT characteristics was correlated with grain size: larger grains result in a larger $\Delta\sigma/\sigma_I$ and sharper $\Delta T_{MI}$. Compared with reported values of $T_{MI}$, $\Delta\sigma/\sigma_I$, and $\Delta T_{MI}$, the characteristic MIT values of our VO$_2$ film were much closer to the corresponding values for a single crystal, suggesting that our VO$_2$ film is of very high quality.

### B. Temperature-dependent XRD measurements



The metallic and insulating phases of $VO_2$ have different crystal structures, and are tetragonal and monoclinic, respectively. A structural phase transition also occurs simultaneously with the MIT. Therefore, we can use careful structural studies, such as XRD, to determine the amounts of the metallic and insulating $VO_2$ phases.

Figure 2(a) presents two characteristic XRD spectra of a $VO_2$ (200) film in its insulating and metallic states. The open and filled squares represent $VO_2$ (200) peaks at 331 and 346 K, respectively. The (200) peak in the insulating state occurs at 37.08°, and that for the metallic state occurs at 37.22°. These values agree well with values for a single crystal.[19] Figure 2(b) presents a contour plot of the temperature-dependent XRD $VO_2$ (200) peak intensity during the heating cycle. The figure shows that as temperature increases, the $VO_2$ (200) peak moved to a higher angle. Note that the XRD peak position changed rather sharply at the temperature region between 335 and 341 K. [In the cooling cycle, the peak position changed between 329 and 335 K.]

To estimate the fraction of metallic phase volume ($p^{XRD}$), we analyzed the XRD spectra quantitatively. The filled circles in Fig. 2(c) represent the $VO_2$ (200) peak at 338 K during the heating run. Note that its full width at half maximum is greater than those at 331 and 346 K. Assuming that two distinct phases coexist inside a $VO_2$ film during its MIT, it is reasonable to fit the broad peak as a superposition of insulating and metallic XRD peaks. As demonstrated in Fig. 2(c), we were able to fit the XRD intensity at 338 K very well in terms of the 331 and 346 K XRD data. We repeated this quantitative XRD analysis for each degree of temperature between 320 and 350 K, and obtained the corresponding $p^{XRD}$ values. The filled circles in Fig. 3(a) represent $p^{XRD}$ values obtained from the XRD analysis. Note



that the temperature dependence of $p^{XRD}$ is very closely correlated with the $dc$ transport near $T_{MI}$, indicated by the solid line.

## C. Spectroscopic images obtained from the STM/STS studies

As noted in the introduction, STS is a powerful microscopic tool for investigating inhomogeneity in surface electronic states. Figure 4 presents STM/STS measurements of the VO$_2$ film surface at 334 K during the cooling cycle. As shown in Fig. 4(a), the topography image revealed that the film had a granular nanocrystalline structure with a root-mean-square roughness of 2 nm. A wider area scan revealed the average grain size to be about 50 nm. Figure 4(c) presents the typical $I$-$V$ characteristics of the metal- and insulator-like regions; the solid lines indicate that Region A should exhibit metallic behavior, and the dashed line indicates that Region B should exhibit an insulating response. The inset shows the $dI/dV$ curves obtained numerically using $I$-$V$ measurement data; it clearly shows that Region A is conducting and Region B yields a vanishing conductance near the Fermi level at zero bias with an insulating energy gap of ~ 0.5 eV. This energy gap value is in good agreement with the bulk value of 0.5 eV.[20] To visualize spatial variation in the electronic states of our film, we recorded the STS images using tunneling current values with a fixed bias of −0.4 V, as shown in Fig. 2(b); this technique is commonly used to illustrate electronic inhomogeneity.[3, 4] Comparison of Figs. 4(a) and (b) suggests that the granular structure of our VO$_2$ film affects the spatial variation in its electronic states.

To observe the progressive evolution of the inhomogeneity in the surface electronic states of our VO$_2$ film, we obtained STS images at numerous temperatures around the MIT.



Figure 5 presents several STS images at decreasing temperatures ranging from 365–300 K. In the spectroscopic images, the black areas represent metallic regions, where the tunneling current becomes smaller than the threshold current level of −5 pA under a sample bias of −0.4 V; correspondingly, the white areas represent insulating regions, where the tunneling current becomes larger than the threshold current level. The fraction of the metallic phase decreased very slowly with the temperature. At 365 K, most of the surface area exhibited metallic behavior, except for small randomly distributed insulating islands. As the temperature decreased, insulating regions appeared and began to grow. Between 339 and 330 K, clusters along the metallic regions appeared to become disconnected, which would constitute real-space observation of percolation. At 300 K, most of the surface area was insulating, except for small randomly distributed metallic islands. Using these STS images, we measured the surface areas of metallic and insulating regions, and estimated the fraction of the metallic region ($p^{STS}$). The filled diamonds in Fig. 3(b) represents temperature-dependent values of $p^{STS}$. Compared to $p^{XRD}$ and the $dc$ transport, $p^{STS}$ revealed a change in MIT over a much wider temperature region.

## IV. DISCUSSION

### A. Application to the percolation model

This section applies the percolation model, which successfully describes a two-phase system across a phase transition, to obtain further insight into the considerable differences



between the $p^{XRD}$ and $p^{STS}$ values. According to the percolation model, σ should have

power law dependence as a function of the metallic volume fraction:

$$\sigma \sim (p - p_c)^t. \qquad (1)$$

where $p$ is a variable representing the metallic fraction, $p_c$ is the percolation threshold of $p$, and $t$ is the conductivity exponent. It is well established that the universal values of $p_c$ and $t$ are 0.45 and 1.4, respectively, for a two-dimensional (2D) percolation conduction model, and 0.15–0.17 and 2.0, respectively, for a three-dimensional (3D) percolation conduction model.[21, 22]

First, we used $p^{XRD}$ values as $p$. As shown in Fig. 6(a), the σ versus $p^{XRD}$ plots for two

different thermal cycles overlapped quite well. The solid line in Fig. 6(a) represents the line

fitted with the percolation model, which agrees fairly well with the experimental data. To

obtain the conductivity exponent, we plotted log (σ) versus log ($p^{XRD} - p^{XRD}_c$), as shown in

the inset of Fig. 6(a). [During actual fitting, we applied the percolation model fitting for a

finite size system.[23]] Based on this, we estimated $t = 2.0 \pm 0.1$ and $p^{XRD}_c = 0.16 \pm 0.01$. These

values are in excellent agreement with the universal values in the 3D percolation models.

These results demonstrate that the MIT transition in our VO$_2$ film occurred during the

percolation process, in which the metallic phase increased in a percolative manner to form

conducting filaments that carry electrical current. Therefore, $p^{XRD}$ should be a good

physical quantity, successfully describing the MIT inside our VO$_2$ film.

Next, we examined what would happen if we used $p^{STS}$ values as $p$. In Fig. 6(b), the open and filled circles represent experimental data during cooling and heating runs, respectively. The solid and dotted lines are the corresponding fitting curves. The log (σ) versus log ($p^{STS} - p^{STS}_c$) curves, shown in the inset in Fig. 6(b), reveal that $t = 0.45 \pm 0.05$. This value is much smaller than the universal values for either the 2D or 3D cases. In addition, the $p^{STS}_c$ values for heating and cooling runs are $0.58 \pm 0.02$ and $0.48 \pm 0.01$, respectively. This difference is unacceptable in the percolation model, where σ should only vary as a function of $p$; the percolation model does not allow two values of $p^{STS}$ for a given σ value. Therefore, $p^{STS}$ cannot describe the percolative evolution of the MIT occurring inside our VO$_2$ film.

### B. Comparison of metallic fractions obtained from experimental methods with differing probing depths



Other experimental studies have investigated the percolative nature of the MIT in $VO_2$ films. Choi *et al*. investigated the mid-infrared properties of a $VO_2$ film near the MIT, and found that the MIT could be described well using the coexistence of metallic and insulating domains.[13] Chang *et al*. extended this research using optical spectroscopy and obtained similar conclusions.[14] In a temperature-dependent photoemission spectroscopy (PES) study of $VO_2$ thin films, Okazaki *et al*. reported hysteretic behavior in the intensity around the Fermi level with temperature across the MIT.[24] Together with our XRD and STS data, we can induce the effects of probing depth when estimating the metallic fractions in $VO_2$ films near the MIT.

Figure 3(a) presents the progressive growth of the metallic fractions in $VO_2$ films, obtained using bulk−sensitive experimental tools, such as XRD and optical spectroscopy. The solid line plots the temperature−dependent σ on a logarithmic scale. The filled circles and open squares represent the fraction of the metallic volume determined using XRD and optical spectroscopy, respectively.[14] The volume fractions estimated using XRD and optical measurements showed very similar temperature dependence, and resembled those of log (σ). The noticeable analogy between log (σ) and $p$ supports the credibility of the scaling law σ ~ $(p − p_c)^t$ based on the percolation model. Note that the probing lengths in XRD and optical measurements exceeded 1000 nm and a few hundred nm (near 0.5 eV), respectively. Therefore, as we expected, the bulk−sensitive experimental tools were able to probe the percolative MIT that occurs inside $VO_2$ films quite well.

By contrast, Figure 3(b) presents the temperature-dependent changes in the metallic content estimated using surface sensitive experimental tools such as PES and STS. The open triangles represent the Fermi level intensities obtained using PES, revealing more gradual temperature dependence, as opposed to the abrupt change in σ. Considering that PES has a probing depth of approximately 1.0 nm at $hv = 21.218$ eV, the fact that it produced different data from the bulk-sensitive tools indicates that inhomogeneities in the surface region within a few lattice constants of $VO_2$ differ from those inside the films. In addition, the filled diamonds in Fig. 3(b) represent the $p^{STS}$ data that we obtained using STS



in this study. Since STS is mainly used to probe electronic states in the top-most layer, the difference between $p^{STS}$ and $p^{XRD}$ can become much larger. These data from surface-sensitive tools suggest that electronic inhomogeneities in the $VO_2$ surface region are quite different from those inside the film and likely exhibit much slower changes near the MIT.

Our STS experiments revealed another aspect of the MIT of our $VO_2$ thin film. If we assume that STS accurately probes the intrinsic metallicity of $VO_2$, it should give a bimodal distribution of the tunneling current at a fixed bias voltage due to the phase separation near MIT. Figure 7 presents the frequency distribution of the tunneling current at –0.4 V. Note that it does not exhibit the bimodal behavior predicted from phase separation, which is a basis of the percolation-type MIT. Instead, it exhibits a slow, gradual change in temperature. All these data suggests that the electronic states at the $VO_2$ film surface differ significantly from those in the interior of the film.

### C. Possible origins

Many semiconductors exhibit numerous surface effects that can affect surface electronic states, and their effects have been investigated systematically.[25] TMOs have a much shorter coherence length of electrons than do semiconductors, so the local state near the surface could be more important. However, it has been difficult to investigate the surface effects of TMO systems, and they are poorly understood.[26] This section discusses why the $VO_2$ surface exhibits such a peculiar phase transition behavior.

Numerous factors could affect surface physical properties, including oxidation states, chemical contaminants, and strain relaxation. First, the oxidation states of vanadium ions at



the surface might differ from +4, possibly due to reduced coordination number, oxygen vacancies, or other defects. Note that there are several binary vanadium oxides, known as the 'Magnéli phases', with a chemical formula $V_nO_{2n-1}$ ($3 \leq n \leq 9$), and these have different oxidation states.[27] These vanadium oxides exhibit a wide range of electronic properties, but no vanadium oxide exhibits MIT at temperatures exceeding the $T_{MI}$ value of $VO_2$: for example, the values for $V_2O_3$ (168 K), $V_6O_{13}$ (145 K), $V_4O_7$ (250 K), $V_5O_9$ (139 K), and $V_6O_{11}$ (177 K) are much lower than for $VO_2$ (341 K).[28] Although the MIT observed using STS appeared to be much broader than those observed using other bulk-sensitive tools, it was nearly completed around 300 K, which is much higher than the $T_{MI}$ values for most other $V_nO_{2n-1}$ phases. Therefore, surface oxidation states might not be able to explain the STS data. Second, chemical contamination is unlikely to affect our results. The STM/STS data were similar for different sample positions and over repeated heating/cooling cycles. In addition, the RHEED pattern also assured that the sample surface was well ordered and did not reveal any degradation features.

Variation in the strain states near the surface might play a very important role in the broad MIT observed using STS. Note that $VO_2$ has strong electron-lattice coupling and considerable electron correlation effects.[16, 19, 29] As noted above, several reports have indicated that the $T_{MI}$ value of epitaxial films can vary from 341–300 K by controlling the strain states.[15-17] Several researchers have also reported that in $VO_2$ films, MIT occurs less abruptly with decreased film thickness.[17, 18, 30, 31] A small perturbation in the strain state near the surface might be enough to nucleate the other electronic phase at a temperature far from the bulk $T_{MI}$. These nucleation sites would be propagated through the $VO_2$ film, so



they could allow the surface MIT to become broader than that inside the film. This interpretation is consistent with the many reported thickness-dependent MITs in $VO_2$ film.[17, 18, 30, 31] It would also explain the observed wide MIT and the considerable hysteretic difference of $T_{MI}$ in $\sigma(T)$ in our $VO_2$ film, as shown in Fig. 1. Since $VO_2$ has very strong electron-lattice interactions, the broad MIT near the surface might result from variation in the strain states.

The effect of strain on the surface electronic states might also be important in other strongly correlated electron TMO systems, such as $Ca_{1-x}Sr_xVO_3$,[32-35] $NdNiO_3$,[36] and CMR manganates.[3, 4] For example, PES has indicated that $CaVO_3$, which is close to the MIT boundary, has a much smaller spectral weight at the Fermi level than does $SrVO_3$.[32] The origin of the large PES spectral difference is controversial.[33-35] Recent PES spectra, obtained at several photoelectron energies, have revealed that the stronger structural ($GdFeO_3$-type) distortion on a $CaVO_3$ versus a $SrVO_3$ surface can cause a stronger dependence of the observed spectral weight for a probing length.[34, 35] Kozuka *et al*. reported that the interface electronic properties of $NdNiO_3$ are distinct from those of the interior and do not undergo bulk MIT.[36] Electronic characterization of the junctions formed between $NdNiO_3$ films and Nb-doped $SrTiO_3$ substrates reveals that $NdNiO_3$ MIT is strongly suppressed at the interface, where the $SrTiO_3$ substrate effectively clamps the $NdNiO_3$ to prevent structural distortion. In addition, it is well-known that the CMR manganates have very strong electron-lattice interactions.[37] Fath *et al*. investigated the spatially inhomogeneous MIT in $La_{1-x}Ca_xMnO_3$ thin films using STS and found that some regions remained insulating even to low temperatures, far into the bulk FM state;[3] this result is



similar to our observations in Fig. 5. All of these examples indicate that the strain distribution near the TMO sample surface could significantly influence the surface electronic state. Such effects would be very important to TMO systems with strong electron-lattice interactions. Further systematic investigations are required to clarify such surface effects.

## V. SUMMARY

In summary, we investigated the electronic inhomogeneity near the first-order metal-insulator transition of a $VO_2$ thin film. Using scanning tunneling spectroscopy (STS) and X-ray diffraction (XRD) measurements, we showed that both metallic and insulating phases likely coexist near the transition temperature. However, the surface-sensitive STS data revealed a much more gradual change in the metallic fraction than the bulk-sensitive XRD data. We found that the XRD data was consistent with the *dc* conductivity change in terms of a 3D percolation model. However, the metallic fraction from the STS data deviated from the percolation predictions, suggesting that the surface electronic state of $VO_2$ film differs from that of the interior. We examined several possible causes for this discrepancy, and found that variation in the strain states near the film surface might play an important role. Our results highlight the importance of using surface-sensitive tools with great care when investigating the surface electronic states of transition metal oxides with strong electron-lattice interactions.

## ACKNOWLEDGMENT



This study was supported financially by Creative Research Initiatives (Functionally Integrated Oxide Heterostucture) from the Ministry of Science and Technology (MOST)/Korea Science and Engineering Foundation (KOSEF)/Center for Strongly Correlated Material Research (CSCMR).

## References


[1] P. A. Cox, *Transition Metal Oxides* (Oxford Science Publications, Oxford, 1992).

[2] M. Imada, A. Fujimor, and Y. Tokura, Rev. Mod. Phys. **70**, 1039 (1998).

[3] M. Fäth, S. Friesem, A. A. Menovsky, Y. Tomioka, J. Aarts, and J. A. Mydosh, Science **285**, 1540 (1999).

[4] T. Becker, C. Streng, Y. Luo, V. Moshnyaga, B. Damaschke, N. Shannon, and K. Samwer, Phys. Rev. Lett. **89**, 237203 (2002).

[5] C. Renner, G. Aeppli, B.-G. Kim, Y.-A. Soh, and S.-W. Cheong, Nature **416**, 518 (2002).

[6] E. Dagotto, Science **309**, 257 (2005).

[7] C. Howald, P. Fournier, and A. Kapitulnik, Phys. Rev. B **64**, 100504(R) (2001).

[8] K. M. Lang, V. Madhavan, J. E. Hoffman, E. W. Hudson, H. Eisaki, S. Uchida, and J. C. Davis, Nature **415**, 412 (2002).

[9] S. H. Pan, J. P. O'Neal, R. L. Badzey, C. Chamon, H. Ding, J. R. Engelbrecht, Z. Wang, H. Eisaki, S. Uchida, A. K. Gupta, K. W. Ng, E. W. Hudson, K. M. Lang, and J. C. Davis, Nature **413**, 282 (2001).





[10] T. Hanaguri, C. Lipien, Y. Kohsaka, D.-H. Lee, M. Azuma, M. Takano, H. Takagi, and J. C. Davis, Nature **430**, 1001 (2004).

[11] J. B. MacChesney and H. J. Guggenheim, J. Phys. Chem. Solids **30**, 225 (1969).

[12] D. Kucharczyk and T. Niklewski, J. Appl. Cryst. **12**, 370 (1979).

[13] H. S. Choi, J. S. Ahn, J. H. Jung, T. W. Noh, and D. H. Kim, Phys. Rev. B **54**, 4621 (1996).

[14] Y. J. Chang, C. H. Koo, J. S. Yang, Y. S. Kim, D. H. Kim, J. S. Lee, T. W. Noh, H.-T. Kim, and B. G. Chae, Thin Solid Films **486**, 46 (2005).

[15] K. Nagashima, T. Yanagida, H. Tanaka, and T. Kawai, J. Appl. Phys. **100**, 063714 (2006).

[16] Y. Muraoka and Z. Hiroi, Appl. Phys. Lett. **80**, 583 (2002).

[17] K. Nagashima, T. Yanagida, H. Tanaka, and T. Kawai, Phys. Rev. B **74**, 172106 (2006).

[18] D. Brassard, S. Fourmaux, M. Jean-Jacques, J. C. Kieffer, and M. A. E. Khakani, Appl. Phys. Lett. **87**, 051910 (2005).

[19] M. Marezio, D. B. McWhan, J. P. Remeika, and P. D. Dernier, Phys. Rev. B **5**, 2541 (1972).

[20] H. W. Verleur, A. S. Barker, and C. N. Berglund, Phys. Rev. **172**, 788 (1968).

[21] M. Sahimi, *Application of Percolation Theory* (Taylor & Francis, London, 1994).

[22] R. Zallen, *The Physics of Amorphous Solids* (Wiley, New York, 1983).

[23] The scaling form of $\sigma \sim (p - p_c)^t$ should be valid when the insulating region is $\sigma = 0$: namely, when the $\sigma$ ratio between the metal and the insulator is infinite.    In the case of


finite $\sigma$, however, the scaling ansatz of the $\sigma$ of a two-component mixture with $\sigma_a$ and $\sigma_b$ ($\sigma_a > \sigma_b$) usually satisfies $\sigma \sim h^u \, g(\varepsilon / h^\eta)$, where $h = \sigma_b/\sigma_a$, $\varepsilon = (\phi - \phi_c)/\phi_c$, $u = t\eta$, and $g(x)$ approaches zero as $g(x) \sim x^t$ for very small x. In our VO$_2$ film, the order of $h$ is about $10^{-4}$. Therefore the fitting form of $\sigma \sim (p - p_c)^t$ is valid for our system since the ansatz approaches the form $\sigma \sim (\phi - \phi_c)^t$ in this region. Refer to D. C. Hong, H. E. Stanley, A. Coniglio, and A. Bunde, Phys. Rev. B **33**, 4564 (1986). .


[24] K. Okazaki, H. Wadati, A. Fujimori, M. Onoda, Y. Muraoka, and Z. Hiroi, Phys. Rev. B **69**, 165104 (2004).

[25] W. Mönch, *Semiconductor Surfaces and Interfaces* (Springer, Berlin, 2001).

[26] P. A. Cox and V. E. Henrich, *The Surface Science of Metal Oxides* (Cambridge University Press, Cambridge, England, 1994).

[27] S. Kachi, K. Kosuge, and H. Okinaka, J. Solid State Chem. **6**, 258 (1973).

[28] J. B. Goodenough, Prog. Solid State Chem. **5**, 145 (1971).

[29] A. Cavalleri, C. Tóth, C. W. Siders, J. A. Squier, F. Ráksi, P. Forget, and J. C. Kieffer, Phys. Rev. Lett. **87**, 237401 (2001).

[30] H. K. Kim, H. You, R. P. Chiarello, H. L. M. Chang, T. J. Zhang, and D. J. Lam, Phys. Rev. B **47**, 12900 (1993).

[31] J. Y. Suh, R. Lopez, L. C. Feldman, and R. F. Haglund, Jr., J. Appl. Phys. **96**, 1209 (2004).

[32] A. Fujimori, I. Hase, H. Namatame, Y. Fujishima, Y. Tokura, H. Eisaki, S. Uchida, K. Takegahara, and F. M. F. de Groot, Phys. Rev. Lett. **69**, 1796 (1992).

[33] K. Maiti, P. Mahadevan, and D. D. Sarma, Phys. Rev. Lett. **80**, 2885 (1998).





[34] A. Sekiyama, H. Fujiwara, S. Imada, S. Suga, H. Eisaki, S. I. Uchida, K. Takegahara, H. Harima, Y. Saitoh, I. A. Nekrasov, G. Keller, D. E. Kondakov, A. V. Kozhevnikov, Th. Pruschke, K. Held, D. Vollhardt, and V. I. Anisimov, Phys. Rev. Lett. **93**, 156402 (2004).

[35] R. Eguchi, T. Kiss, S. Tsuda, T. Shimojima, T. Mizokami, T. Yokoya, A. Chainani, S. Shin, I. H. Inoue, T. Togashi, S. Watanabe, C. Q. Zhang, C. T. Chen, M. Arita, K. Shimada, H. Namatame, and M. Taniguchi, Phys. Rev. Lett. **96**, 076402 (2006).

[36] Y. Kozuka, T. Susaki, and H. Y. Hwang, Appl. Phys. Lett. **88**, 142111 (2006).

[37] A. J. Millis, Nature **392**, 147 (1998).


**FIGURES**



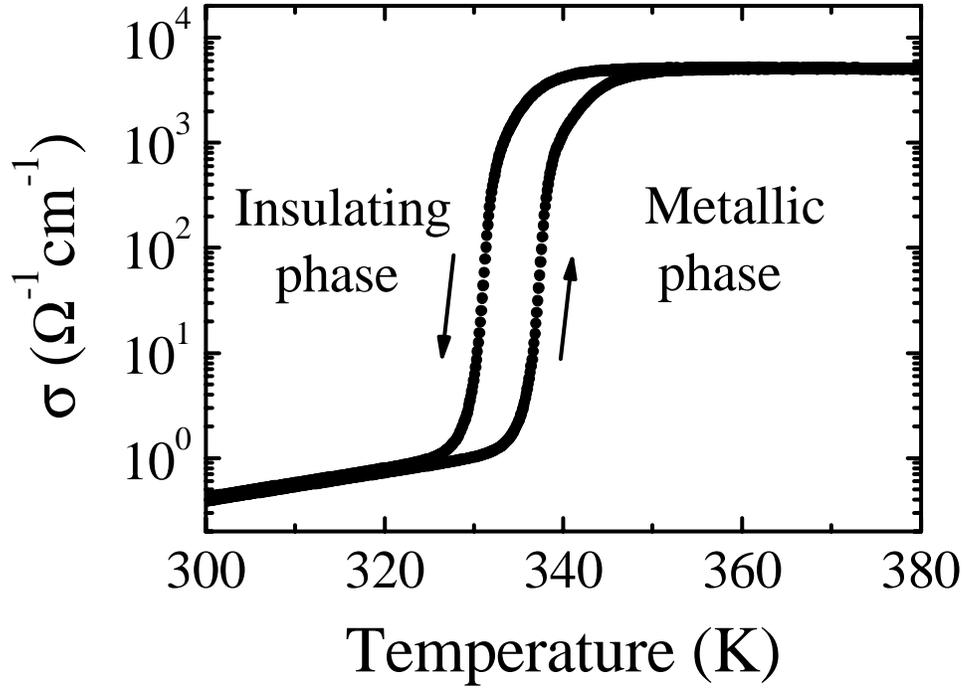

FIG. 1. Electrical conductivity of ($\sigma$) a 70-nm-thick VO$_2$ thin film on Al$_2$O$_3$ (012) substrate; $\sigma$ changed sharply and clearly demonstrated hysteresis. The arrows indicate the direction of the sweeping temperature.



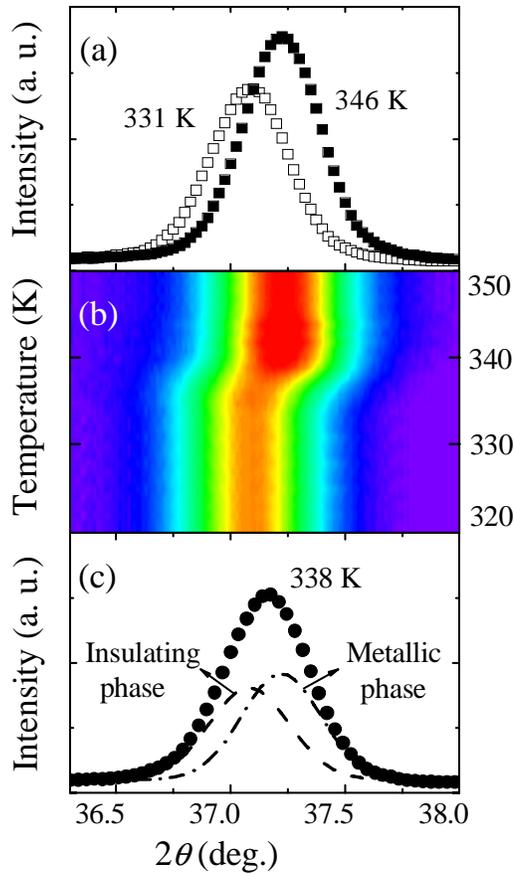

Fig. 2. (Color online) (a) X-ray diffraction (XRD) patterns of the VO$_2$ (200) peak measured at 331 K (open squares) and 346 K (filled squares) during the heating run. (b) A contour plot of the temperature-dependent XRD VO$_2$ (200) peak intensity, obtained while increasing the temperature; the film peak shifted at around 338 K, while it did not move below 331 K or above 346 K, except for negligible thermal expansion. (c) Fitting of the XRD pattern at 338 K (filled circles) as the superposition of the peaks obtained from the insulating (short dashed line) and metallic (dashed line) components.



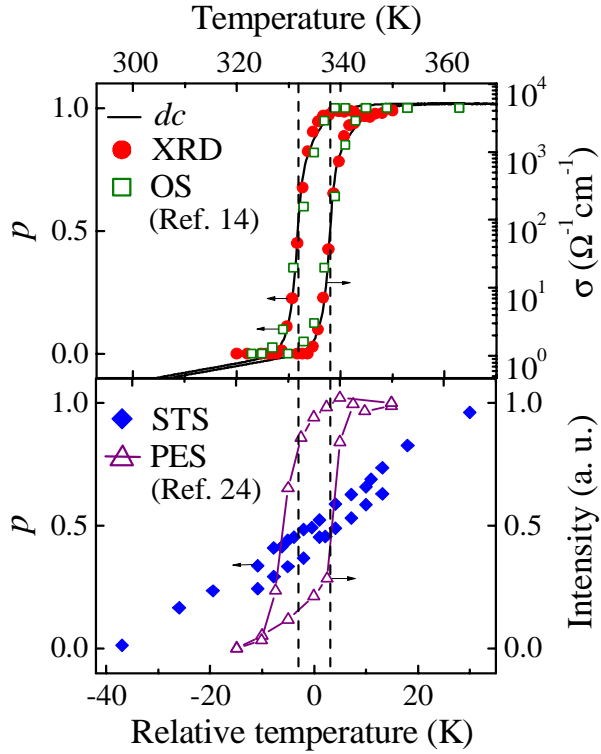

Fig. 3. (Color online) (a) Temperature dependence of the metallic fraction ($p$) extracted from XRD (filled circles) and optical spectroscopy (OS, open squares) measurements [the OS data were adapted from Ref. 14.]. The solid line representing the $\sigma$ data reveals well-matched temperature dependence for both XRD and OS data. (b) Temperature dependence of $p$ estimated from scanning tunneling spectroscopy (STS, filled diamonds) measurements. The spectral intensity was around the Fermi level, measured using photoemission spectroscopy (PES, open triangles) [data adapted from Ref. 24.].



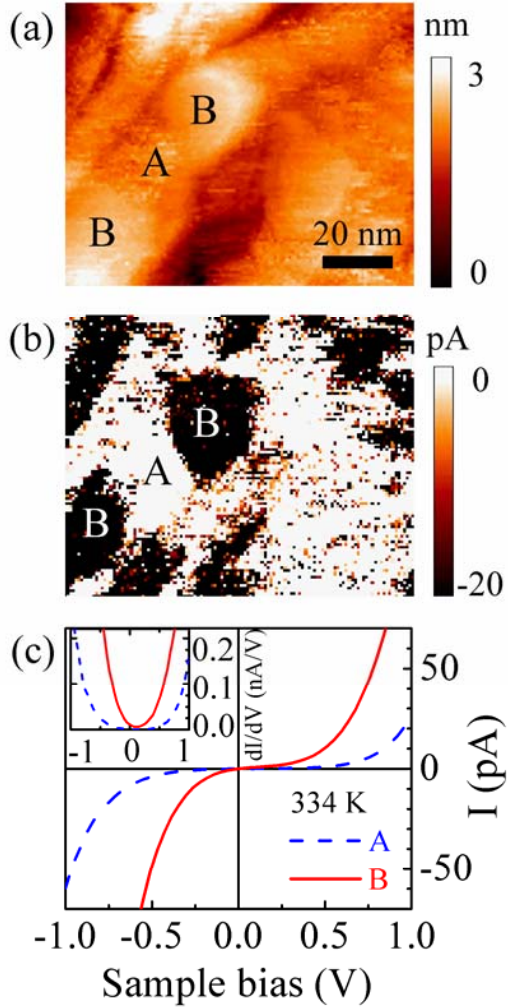

Fig. 4. (Color online) The (a) topography and (b) tunneling current map of the VO$_2$ film, measured simultaneously at 334 K. (c) The *I-V* characteristics curves of regions A and B as indicated in (a) and (b). The inset presents the corresponding tunneling conductance (*dI/dV*) curves. The curve in Region A exhibits a characteristic insulating behavior with a finite energy gap of approximately 0.5 eV.



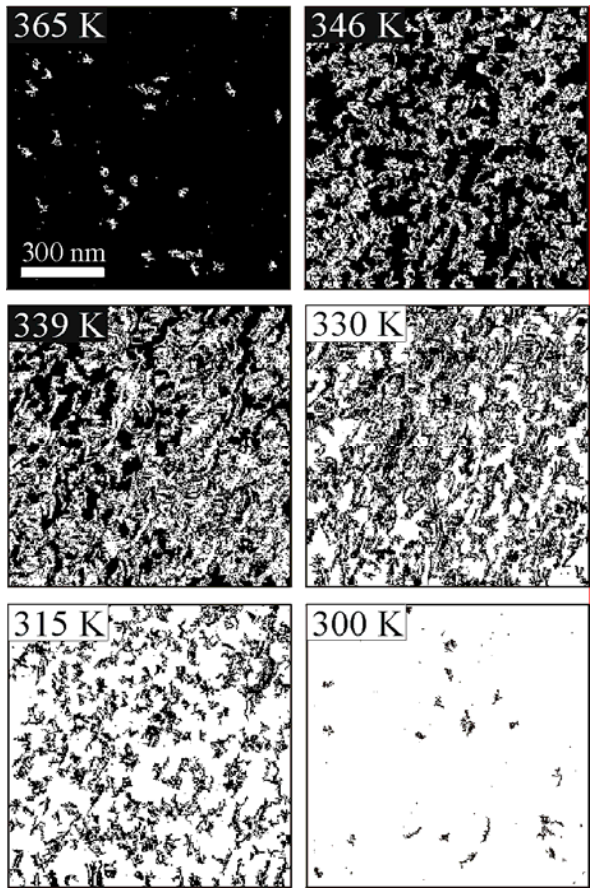

Fig. 5. Spectroscopic images, $1 \times 1 \ \mu m^2$, of the local electronic structure of a $VO_2$ thin film measured at a fixed bias voltage of −0.4 V during cooling from 380 K. Metallic (black) and insulating (white) regions coexisted from 346–315 K. As the temperature decreased, the total area of the metallic regions decreased gradually. At 300 K, almost all areas became insulating regions, except a small portion of the metallic islands.



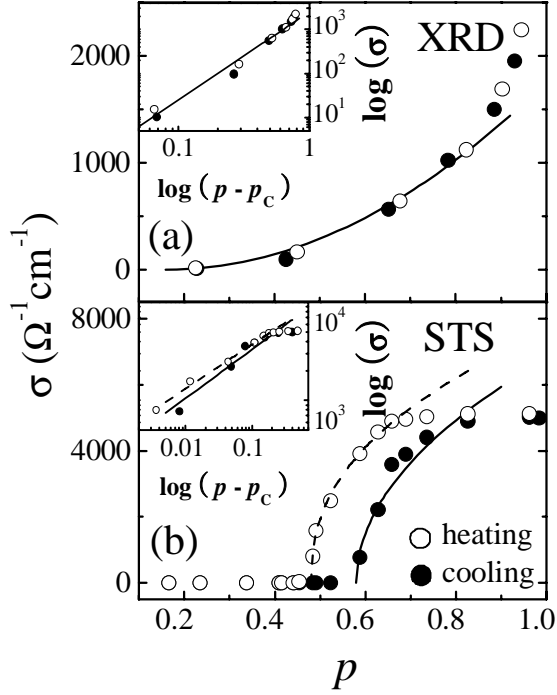

Fig. 6. (a) The $p^{XRD}$ dependence of σ, where the fraction of the metallic volume ($p^{XRD}$) was obtained from XRD analysis. The solid line fits data points to the percolation model; the inset a log-log plot of σ versus ($p^{XRD} - p^{XRD}_c$). The conductivity exponent ($t$) was estimated at 2.0±0.1, which is in good agreement with the universal value of the three-dimensional percolation model. (b) The $p^{STS}$ dependence of σ, where the fraction of the metallic area at the surface region ($p^{STS}$) was estimated from our STS studies. The solid and dashed lines are the fitting curves to each heating and cooling run, respectively. The inset is a log-log plot of σ versus ($p^{STS} - p^{STS}_c$), with an estimated $t = 0.45$±0.05.



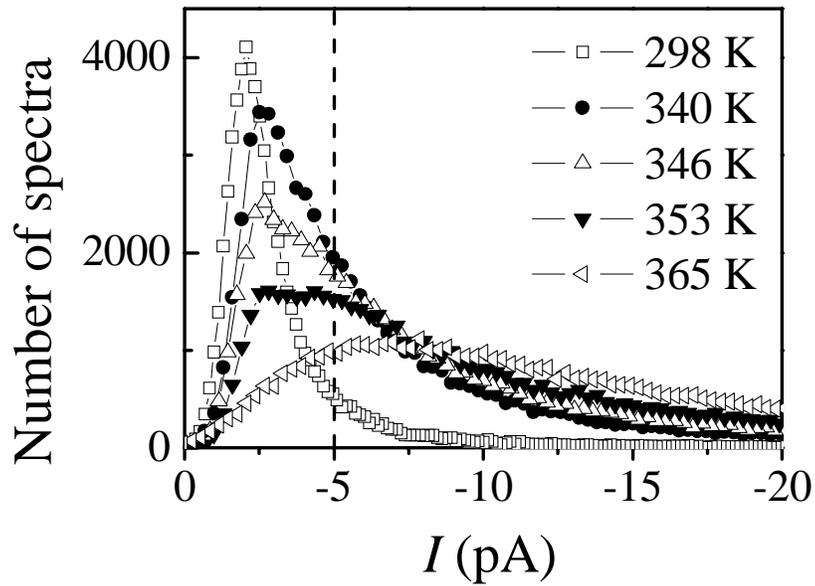

Fig. 7. Frequency distribution of the tunneling current at –0.4 V, obtained from thousands of $I(V)$ scans measured at individual points over large areas at each temperature. The vertical dashed line indicates the threshold for dividing the $I(V)$ scans into metallic or insulating responses in the spectroscopic images shown in Fig. 5.